\documentclass[prx,floatfix,amsmath,amssymb,superscriptaddress,twocolumn]{revtex4}
\usepackage{graphicx}
\usepackage{epstopdf}
\newcommand{\be}{\begin{equation}}
\newcommand{\ee}{\end{equation}}
\usepackage{amsmath}
\usepackage{amsfonts}
\usepackage{bm}
\usepackage{color}
\usepackage{subfigure}
\usepackage{amsthm}
\usepackage[latin1]{inputenc}
\usepackage{tikz}
\usepackage{comment}
\usepackage{xcolor}

\usepackage{soul}

\newcommand{\bra}[1]{\left\langle #1 \right|}
\newcommand{\ket}[1]{\left|#1\right\rangle}
\newcommand{\braket}[2]{\left\langle#1 |  #2\right\rangle}

\newcommand{\Tr}{\textrm{Tr}}

\newtheorem{defi}{Definition}

\newtheorem{myrule}{Rule}

\begin{document}
\title{Ground state entanglement constrains low-energy excitations}
\author{Isaac H. Kim}
\affiliation{Perimeter Institute of Theoretical Physics, Waterloo ON N2L 2Y5, Canada}
\affiliation{Institute for Quantum Computing, University of Waterloo, Waterloo ON N2L 3G1, Canada}
\author{Benjamin J. Brown}
\affiliation{Quantum Optics and Laser Science, Blackett Laboratory, Imperial College London, Prince Consort Road, London, SW7 2AZ, United Kingdom.}

\date{\today}
\begin{abstract}
For a general quantum many-body system, we show that its ground-state entanglement imposes a fundamental constraint on the low-energy excitations. For two-dimensional systems, our result implies that any system that supports anyons must have a nonvanishing topological entanglement entropy. We demonstrate the generality of this argument by applying it to three-dimensional quantum many-body systems, and showing that there is a pair of ground state topological invariants that are associated to their physical boundaries. From the pair, one can determine whether the given boundary can or cannot absorb point-like or line-like excitations.
\end{abstract}

\maketitle
\section{Introduction}
The exotic features of topological phases of matter such as fractional statistics~\cite{Arovas1984}, and genus-dependent ground-state degeneracy~\cite{Wen1990}, are intimately linked to their long-range ground-state entanglement. Indeed, by calculating {\em topological entanglement entropy}~\cite{Hamma2005a, Levin2006, Kitaev2006}, we can extract data of the emergent topological quantum field theory for a given Hamiltonian. The study of such systems is not only of significant fundamental interest, but topological systems also offer a promising route towards quantum information processing in an intrinsically fault-tolerant manner~\cite{Kitaev2003}.

While two-dimensional topological phases are well understood, their three-dimensional~\cite{Hamma2005b, Levin2005a, Walker2011, Vishwanath2013, Metlitski2013}, and higher-dimensional~\cite{Hastings2015} counterparts remain a largely unexplored area of research~\cite{Keyserlingk2012, Bullivant2015}. Of recent interest are the boundaries of three-dimensional phases. Specifically, it has been shown that two-dimensional chiral topological phases, e.g., the semion model, exist on the boundaries of certain three-dimensional phases~\cite{Keyserlingk2012}. Conversely, in some topological phases, we have boundaries that condense particle-like or line-like excitations, which are well studied in two dimensions~\cite{Bravyi1998, Kitaev2012, Levin2013, Barkeshli2013}. Here we develop tools to probe the boundaries of three-dimensional topologically ordered phases using on entropic quantities.

The topological entanglement entropy, $\gamma$, is the constant correction term of the entanglement entropy formula for the ground state of a system
\begin{equation}
S(\rho_A) = \alpha l - n\gamma + \cdots,\label{eq:TEE}
\end{equation}
where $S(\rho_A) = -\Tr(\rho_A \log \rho_A)$ is the von Neumann entropy of subsystem $A$, constant $\alpha$ depends on the microscopic details of the system, $l$ is the length of the boundary that separates subsystem $A$ from its complement and $n$ is the number of disconnected components of the boundary~\cite{Levin2006,Kitaev2006}. Assuming that the low-energy physics is described by a topological quantum field theory, $\gamma$ reveals information about the underlying field theory~\cite{Levin2006, Kitaev2006}, as well as the data of individual anyonic quasiparticles~\cite{Kitaev2006, Dong2008, Brown2013}, and their braiding statistics~\cite{Zhang2011}.

In this Manuscript, we prove a no-go theorem that illuminates the excitation structure of a system without using any prior assumptions about an underlying topological quantum field theory. Instead, we make assumptions only about the support of creation operators of its quasiparticle excitations. Our results extend existing theorems that give conditions for which topological ground-state degeneracy can or cannot be present~\cite{Hastings2004a,Kim2013}. Novel to this work is that our theorem constrains low-energy excitations, not the ground state degeneracy.

More precisely, we prove that the low-energy excitations of a local gapped Hamiltonian are topologically trivial in the case that constant term $\gamma$ vanishes. We show this by proving the following expression
\begin{equation}
\| UV \ket{\psi_0} -VU \ket{\psi_0} \|\leq  O(\gamma^{\frac{1}{2}}),     \label{eq:main_result}
\end{equation}
for ground state $\ket{\psi_0}$ where $\| \ket{\psi}\|=\braket{\psi}{\psi} $ is the norm of the vector. Unitary operator $U$ creates excitations from the vacuum, and $V$ represents a unitary process of (i) creating particles, (ii) performing some non-trivial monodromy operation with a quasiparticle created with $U$, and (iii) annihilating the particles created by $V$.

The result of Eq.\ref{eq:main_result} may seem unsurprising in view of rigorously studied two-dimensional(2D) topological phases~\cite{Kitaev2003,Levin2005}. However, the novelty of our method is that we obtain this result without making any assumptions that depend on the microscopic details of the Hamiltonian. We only assume that we can perform a monodromy operation between particles using operators $U$ and $V$. This generality enables us to perform a similar analysis in more complicated settings which in turn allow us to find new topological invariants.

Another point worth noting is that our method can be easily extended to higher dimensional systems.  Indeed, we explicitly demonstrate the power of our framework by proving that certain linear combinations of entanglement entropies cannot vanish on the boundary of certain three-dimensional(3D) topologically ordered systems that support topological excitations. Specifically, we find a pair of topological invariants that are defined on the boundary, each of which represent the long-range entanglement associated to the point-like and line-like excitations. If the invariant for the point-like excitations is zero, all point-like excitations can be condensed at the boundary. Similarly, if the invariant for the line-like excitations is zero, all such excitations can be condensed at the boundary. We give evidence that these numbers are universal by explicit analytical calculation using the  boundaries of the 3D toric code~\cite{Hamma2005b}. Moreover, we expect these diagonistics to be useful for analyzing Walker-Wang models~\cite{Walker2011,Keyserlingk2012}, as we are able to prove that the invariants must attain a nonzero value for these models. 
This is surprising since previous analyses have shown that bulk topological entanglement entropy give null results for the modular variants of the Walker-Wang models~\cite{Keyserlingk2012, Bullivant2015}.

 Further, our results extend the work of Grover~{\it et al.}~\cite{Grover2011}, where they seek entropic topological invariants in higher-dimensional phases. In their work they show that there is only one invariant in the {\em bulk} of three-dimensional topologically ordered systems. Our results show that the entanglement structure at the boundary of a topological phase can potentially be richer than that of the bulk, as we find two distinct diagnostics that provide information about different types of low-energy excitations at the boundary of a model.

The remainder of this Manuscript is structured as follows; In Sec.~\ref{Sec:TwoDimensions} we prove that a vanishing topological entanglement entropy is a sufficient condition to show that a phase is topologically trivial. For clarity, we present the proof together with the explicit example of a two-dimensional phase. In Sec.~\ref{Sec:ThreeDimensions} we modify our proof for the boundaries of three-dimensional systems. We identify two entropic invariants for identifying different particle types. In Sec.~\ref{Sec:ThreeDimensionalToric} we demonstrate our three-dimensional invariants by consideration of the different boundaries of the three-dimensional toric code before giving some concluding remarks. Technical details of calculations made in Sec.~\ref{Sec:ThreeDimensionalToric} are given in App.~\ref{App:EntropyCalculations}.

\section{Two-dimensional topological phases}
\label{Sec:TwoDimensions}


Let us first sketch the proof that $\gamma$ must be non vanishing for a two-dimensional model to give rise to anyonic excitations. We begin by considering the creation of two quasiparticles by a string-like operator $U$. Then we identify a condition on $U$ that ought to be satisfied for any anyon model. This condition, which shall be explained shortly, implies that the action of $U$ on the ground state can be approximated by a unitary operator $U'$ which lies only in the vicinity of the quasiparticles, with an approximation error that scales as $O(\gamma^{\frac{1}{2}})$. We show this using the fact that $V$ has no common support with $U'$ and thus commutes with $V$. The inequality of  Eq.\ref{eq:main_result} follows from this observation.

These arguments make use of the well-known concepts in quantum information theory, and as such, we set the relevant terminology and definitions first. We use two different distance measures between quantum states $\rho$ and $\sigma$, the fidelity, $F(\rho,\sigma) = \|\rho^{\frac{1}{2}}\sigma^{\frac{1}{2}}\|_1$, and the trace distance, $D(\rho, \sigma) = \frac{1}{2}\|\rho-\sigma \|_1$. These two measures can be used interchangeably, due to their well-known relation~\cite{Nielsen2000}:
\begin{equation}
1-F(\rho,\sigma) \leq D(\rho,\sigma) \leq \sqrt{1-F(\rho,\sigma)^2}.\nonumber
\end{equation}

\begin{figure}
\subfigure[Premise\label{fig:U}]{
\includegraphics[width=1.5in]{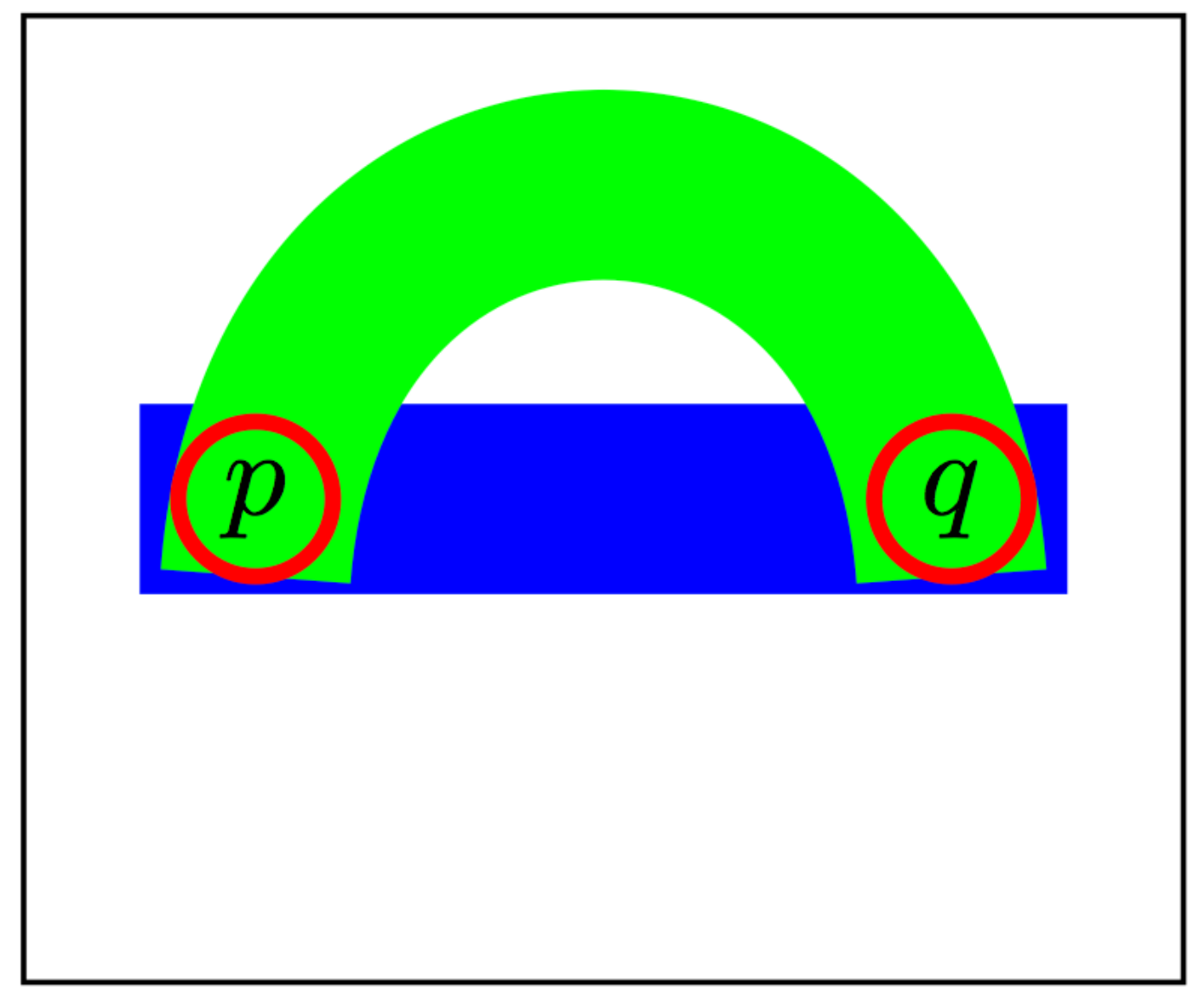}
}
\subfigure[Outcome\label{fig:Uprime}]{
\includegraphics[width=1.5in]{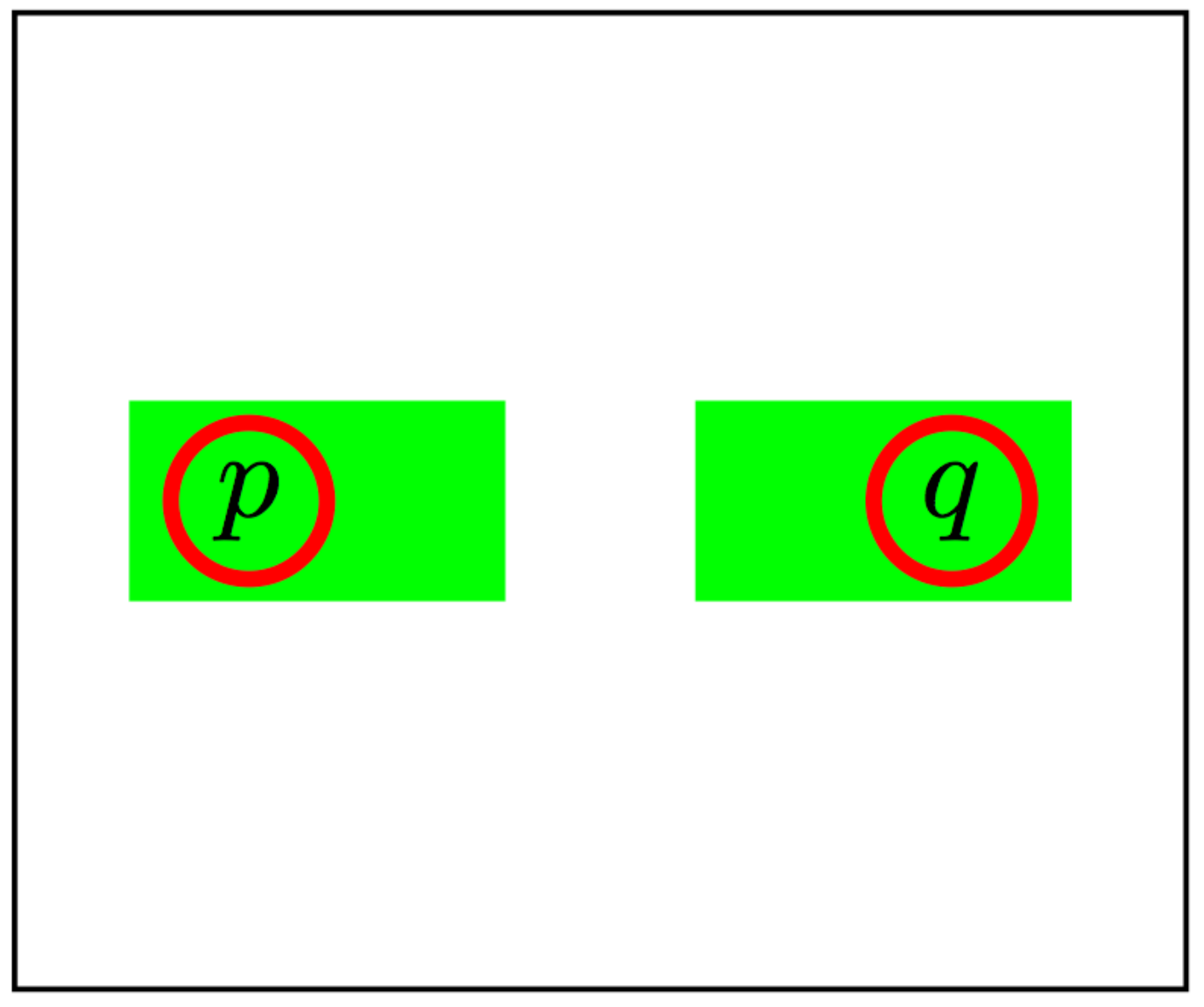}
}
\caption{(Color online) A pictorial representation of the premise and the outcome of the cleaning process. (a) Premise: we assume that the unitary operator $U$(blue) that creates particles $p$ and $q$ can be deformed into $U_{\text{def}}$(green). (b) Outcome: there exists a unitary operator $U'$(green) in the vicinity of the particles such that $\|U\ket{\psi_0} - U'\ket{\psi_0} \|\leq  O(\gamma^{\frac{1}{2}})$. \label{fig:secondstep}}
\end{figure}

Now we go through the details of each steps. Let us begin by stating the most crucial part of the argument, which is pictorially represented in FIG.\ref{fig:secondstep}. To be more specific, consider a pair of quasiparticles created out of the vacuum state $\ket{\psi_0}$ by a string-like unitary operator $U$.  We show that
\begin{equation}
\|U\ket{\psi_0} - U'\ket{\psi_0} \|\leq  O(\gamma^{\frac{1}{2}}), \label{eq:main_result_intermediate}
\end{equation}
for some $U'$ that lies in the vicinity of the particles, if $U$ is \emph{freely deformable}; we say that $U$ is freely deformable if the particles can be created by another string-like unitary operator $U_{\text{def}}$ whose support can be continuously  deformed into that of $U$. This is a natural assumption that is expected to hold for many anyon models. When $\gamma \approx 0$, the above assertion implies that $U\ket{\psi_0} \approx U'\ket{\psi_0}$. In short, the effective support of $U$ is reduced. We refer to such process as the \emph{cleaning process}~\footnote{This method of reducing the support of an operator is of the spirit to the `cleaning lemma', presented for stabilizers in~\cite{Bravyi2008}.}.

The cleaning process relies upon two facts about general quantum states. We first lay out these observations and later explain how they can be applied to anyon models. First, for any two bipartite pure states $\ket{\psi_1}$ and $\ket{\psi_2}$  that have identical density matrices over a subsystem can be mapped onto one another by applying a unitary operation only on the complementary subsystem. Second, there is a condition under which one can check the equivalence of two states from their local subsystems~\cite{Kim2014}. In this paper, we use the second observation to argue that $U\ket{\psi_0}$ and $\ket{\psi_0}$ have the same density matrices over the complement of the support of $U'$ if $\gamma$ is small. Then we use the first observation to argue that there exists a unitary $U'$ which is supported on a smaller region, as explained in FIG.~\ref{fig:secondstep}. We now elaborate on these observations.

The first observation follows from the celebrated Uhlmann's theorem~\cite{Uhlmann1976}, which asserts that $F(\rho,\sigma)$ is equal to the maximum overlap over their purifications:
\begin{equation}
F(\rho,\sigma) = \max_{\ket{\psi_{\sigma}}} |\braket{\psi_{\sigma}}{\psi_{\rho}} |.
\end{equation}
In our context, we envision $\rho$ and $\sigma$ to be the reduced states that are inherited from some bipartite pure states $\ket{\psi_{\rho}}$ and $\ket{\psi_{\sigma}}$. If the fidelity between $\rho$ and $\sigma$ is $1$, the above relation implies that there exists a purification of $\sigma$ that has a unit overlap with $\ket{\psi_{\rho}}$. In particular, it would imply the existence of a unitary operator acting on the complement of the support of $\rho$, such that it maps $\ket{\psi_{\rho}}$ to $\ket{\psi_{\sigma}}$ and vice versa.

\begin{figure}
\includegraphics{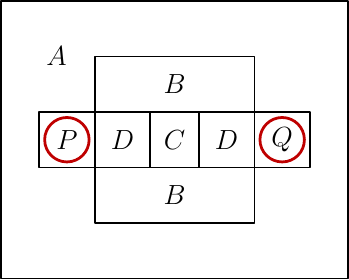}
\caption{(Color online) The division of the system into the relevant subsystems. The particles(red circles) live on $PQ$.\label{fig:consistent_regions} }
\end{figure}

The second observation lies on a recently discovered fact: that two locally equivalent many-body quantum states are globally equivalent under a certain condition. If $\rho_{ABC}$ and $\sigma_{ABC}$ are consistent over $AB$ and $BC$, i.e., $\rho_{AB}=\sigma_{AB}$ and $\rho_{BC}=\sigma_{BC}$, the following inequality holds:
\begin{equation}
D(\rho_{ABC},\sigma_{ABC})^2 \leq I(A:C|B)_{\rho} + I(A:C|B)_{\sigma}, \label{eq:LE_GE}
\end{equation}
where $I(A:C|B)_{\rho} = S(\rho_{AB}) + S(\rho_{BC})- S(\rho_{B}) - S(\rho_{ABC})$ is the conditional mutual information for density matrix $\rho$~\cite{Kim2014}.

So far we have discussed two general facts about quantum states. The natural course is to explain what these facts imply for anyon models. Without loss of generality, let us choose $\rho$ to be the ground state, i.e.,  $\rho = \ket{\psi_0} \! \! \bra{\psi_0}$ and $\sigma$ to be the excited state, i.e., $U\rho U^{\dagger}  = U_{\text{def}} \rho {U_{\text{def}}}^{\dagger}$. We divide the systems into the regions shown in FIG.\ref{fig:consistent_regions}, for reasons that will soon become apparent. It should be noted that $\rho$ and $\sigma$ must have the  same density matrices over $AB$ and $BC$ since $U$ can be deformed to have a support complementary to these regions. Importantly, this implies that we can use Eq.\ref{eq:LE_GE}.

We estimate the right-hand side of Eq.\ref{eq:LE_GE} for the choices we have just made. Quantum entropy obeys  strong subadditivity of entropy~\cite{Lieb1972}, which implies that $I(A:C|B)_{\rho}\leq I(APQ:C|B)_{\rho}$. Recall that entanglement entropy over a region is equal to the entanglement entropy over its complement, if the global state is pure. Therefore, the right-hand-side of Eq.\ref{eq:LE_GE} can be bounded by the sum of $S(\rho_{BC}) + S(\rho_{CD}) - S(\rho_{B}) - S(\rho_{D})$ and $S(\sigma_{BC}) + S(\sigma_{CD}) - S(\sigma_{B}) - S(\sigma_{D})$. Since $U$ can be freely deformed to be supported on the complement of $BCD$, we have that $S(\rho_R) = S(\sigma_R)$ for $R =B,\, D,\, BC $ and $ CD$. We therefore obtain the bound
\begin{eqnarray}
D(\rho_{ABC},\sigma_{ABC})^2 \leq  2 [  S(\rho_{BC}) &+& S(\rho_{CD}) \nonumber
\\ & - &S(\rho_{B}) - S(\rho_{D}) ].
\end{eqnarray}

Having obtained an upper bound for $D(\rho_{ABC},\sigma_{ABC})^2$ that depends only on the ground state $\rho$, it can be evaluated for topologically ordered states using Eq.\ref{eq:TEE}. We arrive at the conclusion that $\rho_{ABC}$ can be approximated by $\sigma_{ABC}$ with an approximation error of $2\gamma^{\frac{1}{2}}$, i.e., $D(\rho_{ABC}, \sigma_{ABC}) \leq 2 \gamma^{\frac{1}{2}}.$ If $\gamma \approx 0$, $\rho_{ABC} \approx \sigma_{ABC}$. By Uhlmann's theorem, this would imply that $U\ket{\psi_0}$ can be mapped into $\ket{\psi_0}$ by applying a unitary operator on the complement of $ABC$, thus proving Eq.\ref{eq:main_result_intermediate}.

Intuitively, this leads to a contradiction if the particle carries a nontrivial topological charge. This is due to the defining characteristics of such particles: that they cannot be created or annihilated locally. We use two simple facts to show this concretely. First, $V\ket{\psi_0}= e^{i\phi} \ket{\psi_0}$. This means that the process $V$ acts trivially on the ground state. Second, $V$ commutes with $U'$. This is due to the fact that the support of $U'$ lies only in the vicinity of the quasiparticles, whereas the support of $V$ can be made to be far away from the quasiparticles. Since the norm is invariant under unitary rotation,
\begin{equation}
\|VU\ket{\psi_0} - VU' \ket{\psi_0} \| = \|U\ket{\psi_0} - U'\ket{\psi_0} \| \leq O(\gamma^{\frac{1}{2}}).
\end{equation}
It should be noted that $VU' \ket{\psi_0}$ is actually equal to $U'V\ket{\psi_0}$ due to the commutation relation. Since $V$ acts trivially on the ground state,
\begin{equation}
\|U'V\ket{\psi_0} - UV\ket{\psi_0} \| = \|U'\ket{\psi_0} -U\ket{\psi_0} \|\leq O(\gamma^{\frac{1}{2}}).
\end{equation}
Applying the triangle inequality to the above two inequalities, we arrive at Eq.\ref{eq:main_result}.

\section{Three-dimensional topological phases}
\label{Sec:ThreeDimensions}

So far we have explained why $\gamma$ must attain a nonzero value if anyons exist in two-dimensional systems; otherwise any
excitation can be created locally from the vacuum. This is an instructive example which demonstrates the fundamental connection between the ground-state entanglement and the properties of the low-energy excitations. This intuition can be extended to systems of higher dimension to probe the nature of different types of quasiparticle excitations. Further, we can develop our intuition to study the boundaries of topological phases, where the physics of a system will change.

Near the boundary, certain topologically nontrivial excitations can be created locally out of the vacuum.  This is because certain boundaries are capable of absorbing, or `condensing' certain topological excitations~\cite{Bravyi1998, Kitaev2012, Levin2013, Barkeshli2013}. As such, the aforementioned argument can be modified accordingly to identify boundaries that condense topological charges. Conversely, it follows from our argument that phases that support topological excitations on their boundaries necessarily have nonzero topological entanglement entropy.

\begin{figure}
\includegraphics[width=\columnwidth]{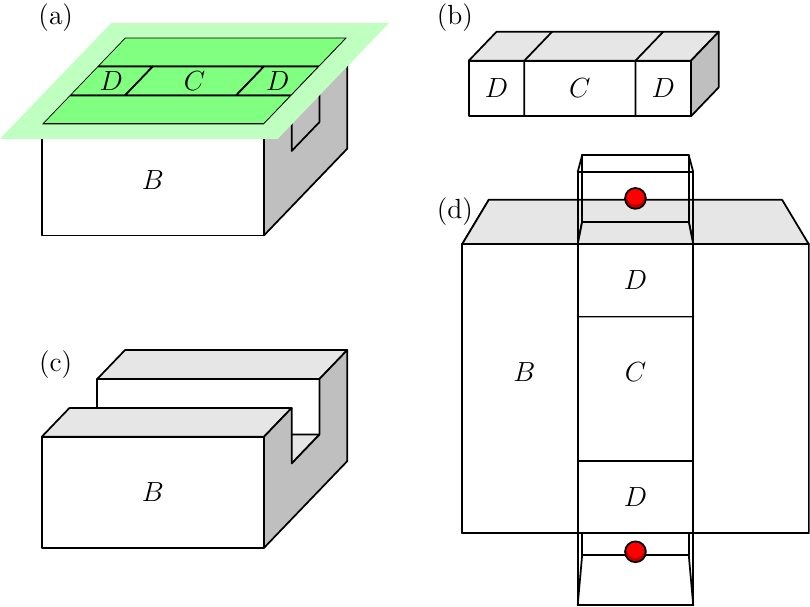}
\caption{(Color online) Regions that define $\gamma_{\text{point}}$. The green surface in (a) is the physical boundary. (b) and (c) show regions $CD$, and $B$, respectively. The red dots in (d) represent the point-like excitations we wish to examine. \label{fig:obstructions_3D_point}}
\end{figure}

Remarkably, in 3D, topological phases can host exotic line-like quasiparticle excitations that carry nontrivial topological charge, as well as point-like excitations. To this end we can construct topological invariants to identify both point-like and line-like topological excitations by consideration of the support of their creation operators.

We give two topological invariants that are applicable to the boundaries of three-dimensional topological phases. The first, the {\em point topological entanglement entropy}, is designed to learn the nature of point-like particles near a boundary. The second, the {\em line topological entanglement entropy}, achieves a null value for boundaries where all line-like excitations are topologically trivial. Our invariants are obtained following an argument similar to that given in the previous Section.

\subsection{Point topological entanglement entropy}

We define the point topological entanglement entropy,  $\gamma_{\text{point}}$, as
\begin{equation}
 \gamma_{\text{point}} = S(\rho_{BC}) + S(\rho_{CD}) - S(\rho_{B}) - S(\rho_{D}), \label{Eqn:GammaPoint}
\end{equation}
where regions $B$, $C$, and $D$ are shown in FIG.\ref{fig:obstructions_3D_point}. Region $A$ is the complementary subsystem of region $BCD$. The regions in FIG.\ref{fig:obstructions_3D_point} are labeled such that they perform analogous roles to the regions with the same labels in FIG.\ref{fig:consistent_regions} in the 2D argument given in the previous Section.  For brevity, we have not shown the regions $P$ and $Q$  we have used in the previous Section. These regions are implicitly included in region $A$ adjacent to the parts of region $D$ where the quasiparticles are created. Drawing this analogy allows us to generalize the 2D argument in a natural way to study point excitations on the boundaries of 3D systems. At a high level, one can imagine creating a pair of quasiparticles by applying a deformable string-like operator $U$ that is supported on subsystem $CD$ that, importantly, includes part of the boundary. If $ \gamma_{\text{point}}$ is small, the action of $U$ on the ground state can be approximated by $U'$ which lies in the vicinity of the quasiparticles. Such $U'$ exists only if the quasiparticles can be created locally near to a boundary. Conversely, if there are any point-like excitations that cannot be created by such $U'$, $ \gamma_{\text{point}}$ cannot vanish.

As we did in the 2D case, let us compare two states, the vacuum state, $\rho$, and an excited state with two point-like excitations, $\sigma$; see FIG.\ref{fig:obstructions_3D_point}(d). By our assumption that $U$ is freely deformable, both states have identical density matrices over $BC$ and $CD$. With $A$ the complement of the regions depicted in FIG.\ref{fig:obstructions_3D_point}(a), the trace distance $D(\rho_{ABC}, \sigma_{ABC})$ is upper bounded by ${\gamma_{\text{point}}}^{1/2}$. If $\gamma_{\text{point}}$ is $0$, Eq.\ref{eq:LE_GE} implies that $\rho$ and $\sigma$ are identical over $ABC$. By invoking Uhlmann's theorem, we conclude that there must exist a unitary operator in the complement of $ABC$ that maps $\rho$ to $\sigma$. Since this region is in the vicinity of the particles, we conclude that the particles can be annihilated or created locally. If there are point-like excitations that cannot be condensed at the boundary, $\gamma_{\text{point}}$ cannot vanish.

\subsection{Line topological entanglement entropy}
\begin{figure}[b]
\includegraphics[width=\columnwidth]{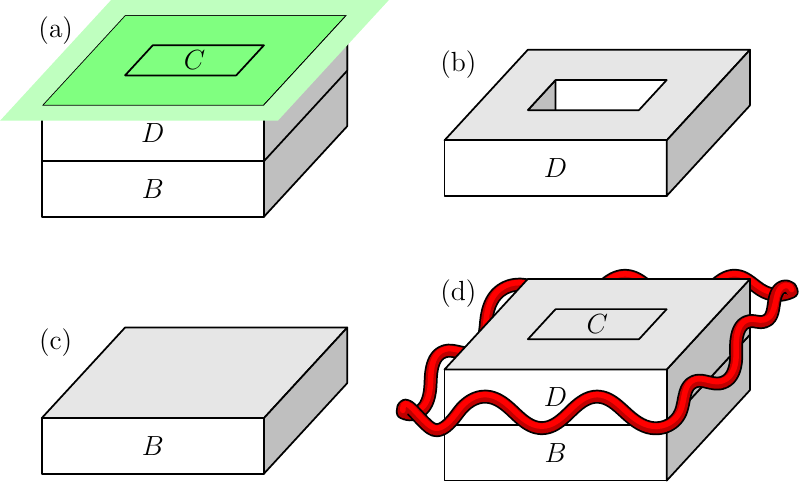}
\caption{(Color online) Regions that define $\gamma_{\text{line}}$. (a) The green surface is the physical boundary. (b) and (c) show individually the annular region $D$ and spherical region $B$, respectively. The red line in (d) represents the line-like excitation we wish to identify.\label{fig:obstructions_3D_line}}
\end{figure}

A similar argument can be carried out for the line-like excitations. We define the {\em line topological entanglement entropy}, $\gamma_{\text{line}}$, by the equation
\begin{equation}
\gamma_{\text{line}} = S(\rho_{BC})+S(\rho_{CD})-S(\rho_{B})-S(\rho_{D}), \label{Eqn:GammaLine}
\end{equation}
where regions $B$, $C$, and $D$ are shown in FIG.\ref{fig:obstructions_3D_line}. Again, subsystem $A$ is the complement of subsystem $BCD$.

Once again our previous argument holds; we imagine creating a line-like excitation by applying a unitary operator $U$ that has nontrivial support on subsystem $CD$, as is shown in FIG.\ref{fig:obstructions_3D_line}(d). If $\gamma_{\text{line}} $ is small, the action of $U$ on the ground state can be approximated by $U'$ which lies in the vicinity of the line-like excitations. Such $U'$ exists only if  either the system has no topologically nontrivial line particles, or if the boundary can absorb all the line-like excitations of the system. As before, it is also true that if there are any line-like excitations that cannot be created by some $U'$, the quantity $\gamma_{\text{line}} $ cannot vanish.

For completeness we explicitly make the argument explaining why the line topological entanglement entropy is a topological invariant. We compare two states, the vacuum state, $\rho$, and an excited state with a loop-like excitation, $\sigma$. By our assumption that $U$ is freely deformable, both states have identical density matrices over $BC$ and $CD$. As we did previously, we denote $A$ as the complement of the regions depicted in FIG.\ref{fig:obstructions_3D_line}(a). The  trace distance $D(\rho_{ABC}, \sigma_{ABC})$ for such regions is upper bounded by ${\gamma_{\text{line}}}^{1/2}$. If $\gamma_{\text{line}}$ is $0$, Eq.\ref{eq:LE_GE} implies that $\rho$ and $\sigma$ are identical over $ABC$. Uhlmann's theorem then implies that there must exist a unitary operator in the complement of $ABC$ that maps these two states. Since this region is a solid torus that surrounds the loop-like excitation, vanishing line topological entanglement entropy implies that the loop-like excitation can be condensed at the boundary.

\subsection{Universality}
In the 2D case, the linear combination was concocted in such a way that the area terms in Eq.\ref{eq:TEE} cancel each other out. Based on a general physical intuition that the leading term is due to the short-range entanglement across the cut, we expect a similar behavior for the regions in FIG.\ref{fig:obstructions_3D_point} and FIG.\ref{fig:obstructions_3D_line}. It should be noted that the physical boundary does not contribute to such short-range entanglement, since the vacuum that lies beyond the physical boundary is not entangled with the medium. Assuming such a behavior indeed holds, one can easily see that the contributions from the short-range entanglement are canceled out.

The remaining term is invariant under smooth deformation of the regions. Therefore, we expect it to be a topological invariant that characterizes the phase. In particular, we have shown that the point(line) topological entanglement entropy becomes $0$ only if all the point-like(line-like) excitations can be condensed at the given boundary. Moreover, our arguments show that we expect positive values for $\gamma_\text{point}$ and $\gamma_\text{line}$ if the studied boundaries support nontrivial point-like or line-like excitations, respectively. This is surprising given the recent results in Refs.~\cite{Keyserlingk2012, Bullivant2015} where it is shown that certain topological phases of matter with topological excitations on the boundary do not give rise to positive topological order parameters when one studies the bulk of the system. In contrast, our argument proves that the point topological entanglement entropy must be nontrivial for modular Walker-Wang models~\cite{Walker2011,Keyserlingk2012}.

We point out that while our diagnostics give to positive values for boundaries where topological excitations are realized, we have not shown that a nonzero value guarantees a system with topological excitations at the boundary. It seems unlikely that one could give such a proof as examples of topologically trivial systems that show nontrivial topological behaviour with respect to certain entropic invariants are known~\cite{Bravyi_Counterexample2012}. To this end, one must be wary when using our entropic invariants, or indeed, any entropic invariants to identify topological order.

\section{Analyzing the three-dimensional toric code}
\label{Sec:ThreeDimensionalToric}
 In Sec.~\ref{Sec:ThreeDimensions} we have introduced two ground-state topological invariants, and we have argued they will give nonzero values for models that give rise to topological excitations on their boundary. In this Section we use the point topological entanglement entropy and the line topological entanglement entropy to examine the different boundaries of the well understood model, the 3D toric code. In particular, we show that our invariants can be used to determine properties of different boundaries with respect to the types of excitations they are able to absorb.

\begin{figure}
\includegraphics{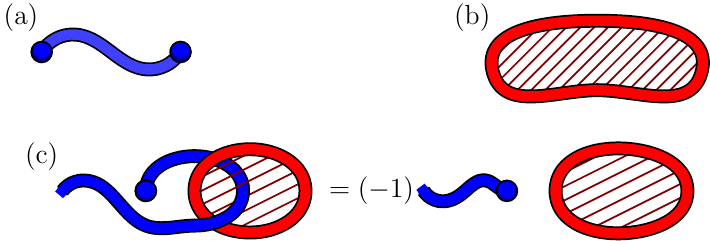}
\caption{(a) Two point-like excitations created at the end points of a string operator. (b) A line-like excitation that is created on the boundary of a membrane operator. (c) A point-like excitation braided through the line-like excitation and returned to its initial position introduces a non-trivial $-1$ phase to the system.\label{ToricExcitations}}
\end{figure}

The 3D toric code~\cite{Hamma2005b} in the bulk has two-types of excitations; one point-like excitation and one line-like excitation, as shown in FIG.~\ref{ToricExcitations}~(a) and~(b) respectively. Point-like excitations are created in pairs at the endpoints of  string-like creation operators, and line-like excitations form closed loops around the boundary of membrane-like creation operators. The model acquires an $e^{\text{i}\pi}$ phase if a point excitation is moved through a closed line excitation and returned to its initial position, as shown in FIG.\ref{ToricExcitations}(c).

The toric code has two types of boundary, a {\em rough boundary} and a {\em smooth boundary}. The 3D boundaries generalize straightforwardly from the 2D case~\cite{Bravyi1998}. Close to a boundary, the excitations of the model change non-trivially. A rough boundary absorbs point-like excitations. Therefore, in the vicinity of a rough boundary, we find only line-like excitations are topologically nontrivial. Conversely, a smooth boundary absorbs line-like excitations. We see that the presented diagnostics can distinguish these different boundaries for the considered example.

\subsection{The von Neumann entropy for subsystems of the three-dimensional toric code}
To employ our topological invariants, we must first find a general formula for the von Neumann entropy of subsystems of the 3D toric code where subsystems may include qubits at either a rough or a smooth boundary.

The bulk entanglement entropy of region $R$ for the 3D toric code~\cite{Castelnovo2008, Grover2011, Keyserlingk2012} is
\begin{equation}
S(\rho_R) =  A_R - n_R \log 2, \label{ToricEnt}
\end{equation}
where $A_R$ is the surface area of the boundary of region $R $, denoted $\partial R$. The term $n_R$ is the number of disjoint connected surfaces, $\partial R_j$, of $\partial R$, such that $\partial R = \partial R_1 \sqcup \partial R_2 \sqcup \dots \sqcup \partial R_n$. 

To calculate $\gamma_{\text{point}} $ and  $\gamma_{\text{line}}$, we generalize Eq.\ref{ToricEnt} for the toric code to regions that include boundary qubits. These calculations are found explicitly using the method of~\cite{Fattal2004} in App.~\ref{App:EntropyCalculations}. To summarize App.~\ref{App:EntropyCalculations}, we find that the topological contribution from boundary component $\partial R_j$ that bounds qubits from a smooth boundary is unchanged. Therefore the boundary component $\partial R_j$ contributes a single unit to the topological term. In constrast, we find that each boundary component, $\partial R_j$, that bounds any qubits of the rough boundary will contribute nothing to the topological term. We therefore arrive at the general formula
\begin{equation}
S(\rho_R) =  A_R - N_R \log 2, \label{Eqn:GeneralFormula}
\end{equation}
with $A_R$ the surface area contribution of the boundary of region $R $, and $N_R$, the number of disjoint boundary components $\partial R_j$ that enclose no qubits from a rough boundary.

\subsection{The smooth boundary of the three-dimensional toric code}

We can apply Eq.\ref{Eqn:GeneralFormula} to find $\gamma_{\text{point}} $ and  $\gamma_{\text{line}} $ for the smooth boundaries of the 3D toric code. For the regions given in FIG.\ref{fig:obstructions_3D_point} we have
\begin{equation}
N_{BC} = 1, \, N_{CD} = 1, \, N_{B} = 1,   \, N_{D} = 2.
\end{equation}
Similarly, for the regions given in FIG.\ref{fig:obstructions_3D_line} we have
\begin{equation}
N_{BC} = 1, \, N_{CD} = 1, \,N_{B} = 1, \, N_{D} = 1.
\end{equation}

Given that the local contributions for the terms in Eq.\ref{Eqn:GammaPoint} and Eq.\ref{Eqn:GammaLine} cancel, we obtain
\begin{equation}
\gamma_{\text{point}} = \log 2,  \quad \gamma_{\text{line}}= 0 ,\label{Eqn:GammaPoint_3DTC}
\end{equation}
at a smooth boundary of the 3D toric code. As predicted, this result is indicative of the existence of topological point particles that cannot be absorbed at the boundary. The negative result for $\gamma_{\text{line}}$ shows that all line-like excitations are absorbed by the smooth boundary.

\subsection{The rough boundary of the three-dimensional toric code}
We finally evaulate $\gamma_{\text{point}} $ and  $\gamma_{\text{line}} $ for the rough boundary of the 3D toric code. We find that $N_R = 0$ for all regions used in Eq.\ref{Eqn:GammaPoint}, as all the disjoint components of the boundaries of the regions in FIG.\ref{fig:obstructions_3D_point} enclose qubits in the rough boundary. Conversely, we have that $N_B = 1$ for region $B$ in FIG.\ref{fig:obstructions_3D_line}, as region $B$ does not touch the boundary. Otherwise we have $N_R = 0$ for all $R \not=  B$ that are used to find $\gamma_\text{line}$ in Eq.\ref{Eqn:GammaLine}. We thus obtain
\begin{equation}
\gamma_{\text{point}} = 0,  \quad \gamma_{\text{line}}= \log 2. \label{Eqn:GammaLine_3DTC}
\end{equation}
Once again, these are the expected results given that the rough boundary absorbs all the point-like excitations of the 3D toric code, but  does not absorb line-like excitations. This result, together with Eq.\ref{Eqn:GammaPoint_3DTC} demonstrates that we can identify boundaries that condense point-like or line-like excitations using our invariants. This is indicated by the null values of $\gamma_{\text{point}}$ or $\gamma_{\text{line}}$.

\section{Conclusion}
 By consideration of the support of quasiparticle creation operators we have shown that we can obtain new entropic invariants for local gapped Hamiltonians using information theoretic arguments. We have used these methods to find two new order parameters for the boundary theories of 3D topological models. We have demonstrated that the proposed measures are effective by studying the boundaries of the 3D toric code. The result we obtain is remarkable given that we cannot distinguish between different excitation types in the bulk of 3D topological phases using entropic diagnostics~\cite{Grover2011}.

 One might consider using the proposed topological invariants to interrogate the structure of more general classes of topologically ordered systems~\cite{Walker2011} with exotic surface theories, where perhaps the bulk topological entanglement contribution is zero~\cite{Keyserlingk2012}. It will be interesting to find a quantitative expression for more general theories of boundary excitations using our methods.  Another class of models of recent interest in this respect are bosonic topological insulators with surface anyon theories~\cite{Vishwanath2013, Metlitski2013}. One might also consider using the present general proof to find new topological invariants for other interesting phases such as fractal topological quantum field theories~\cite{Haah2011,Yoshida2013}.

IK's research at Perimeter Institute is supported in part by the Government of Canada through NSERC and by the Province of Ontario through MRI. BJB is supported by the EPSRC.

\appendix
\section{The von Neumann entropy of the three-dimensional toric code}
\label{App:EntropyCalculations}

\begin{figure}[b]
\includegraphics{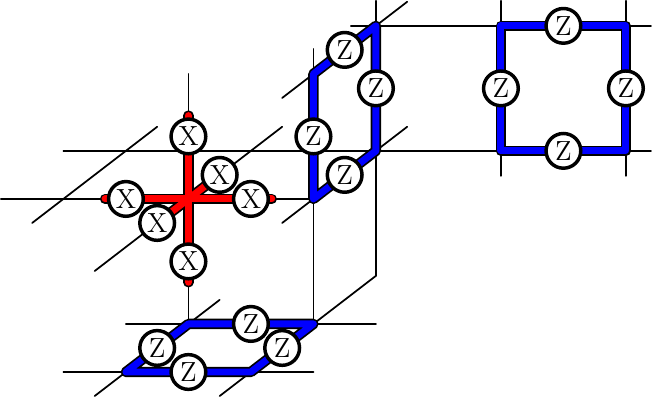}
\caption{A star operator shown in red. The star operator for vertex $v$ supports a Pauli-X operator on each of the edges incident on vertex $v$. Some plaquette operators, shown in blue on different planes. Plaquettes have a Pauli-Z operator on each of the edges that bounds the face of the square lattice. \label{Stabilizers}}
\end{figure}

Here we study the bipartite entanglement between simple regions of the ground state of the 3D toric code lattice. We use the method given in Ref.~\cite{Fattal2004} to find the entanglement entropy of a ball-shaped region in the bulk, and ball-shaped regions that enclose some of the qubits in a rough and a smooth boundary.

The 3D toric code is defined on a square lattice with qubits arranged on its edges. Its degenerate ground space, spanned by basis vectors $\ket{\psi_j}$, is described using the stabilizer formalism~\cite{Gottesman1997}. Specifically, it is described by its (Abelian) stabilizer group, $\mathcal{S} = \left\{ S\in \mathcal{S} : S\ket{\psi_j} = \ket{\psi_j} \forall j \right\} $. The stabilizer group for the 3D toric code contains two types of stabilizers; star and plaquette operators, shown in FIG.\ref{Stabilizers}.

We use the method of Fattal {\it et al.}~\cite{Fattal2004} to find the entanglement entropy between two subsystems, $A$ and $B$, which we briefly summarize. We consider an independent generating set of the stabilizer group with elements $S_j  \in \mathcal{S}$. We write the generators $S_j = S_j^A \otimes S_j^B$, where $S_j^A$ is supported on subsystem $A$ and $S_j^B$ is supported on subsystem $B$. We study the restriction of the generating set of one of the subsystems of interest. Without loss of generality, we consider the restriction of the stabilizer group on subsystem $A$.

The restricted stabilizer generators, $S_j^A$, do not in general commute. The method of Fattal~{\it et al.} looks to find a generating set where each restricted generator either commute with all other restricted generators, or anti commute with only one other restricted generator. Specifically, we look for $2k$ elements of the restricted generating set that satisfies
\begin{equation}
\left\{ S^A_{2j-1}, S^A_{2j}  \right\} = 0,
\end{equation}
for all $1 \le j \le k$. The state described by $\mathcal{S}$ shares $k$ ebits of entanglement between subsystem $A$ and $B$. Generating sets where we are able to count {\em pairs} of anti-commuting operators when restricted to a subsystem are said to be in {\em canonical form}. The result of Fattal {\it et al.} shows that it is always possible to find a generating set in canonical form for any bipartition of the stabilizer group.

We must find a generating set of the stabilizer group of the 3D toric code that is in canonical form under a given bipartition. This enables us to count the ebits shared between two subsystems. Importantly, the generating set is over complete if we include $B_f$ operators for all the faces. This is seen by taking the product of all the plaquette operators corresponding to the faces that bound a cube. This product returns identity, showing an over-complete generating set where eigenvalues of stabilizers are dependent on others.

We choose an independent generating set that includes all plaquette operators that lie parallel to the $xy$ and $yz$ plane, and we only take the plaquette operators parallel to the $xz$ plane in a single plane at some fixed $y$. We are free to choose which plane, and for simplicity we always take this plane to be far away from the region of interest for the entropy calculation. For this reason, for all the calculations we make, it is sufficient to consider the only the plaquette generators parallel to the $xy$ and $yz$ plane.

Similarly, we point out now that we need not account for the logical operators that may appear in the generating set of the stabilizer group. Logical operators can always be deformed away from the regions of interest on the lattice and as such never contribute to the entanglement in any of the bipartitions we study. Moreover, our results are independent on the choice of ground state.

\subsection{The von Neumann entropy of a ball in the bulk}

\begin{figure}[t]
\includegraphics{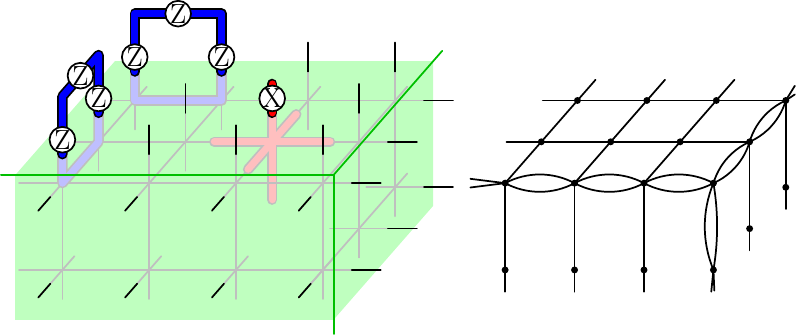}
\caption{\label{Jelly} (Left) The corner of a ball-shaped region, labeled $A$, of the 3D toric code. We show the support of one star and two plaquette operators on region $B$ by Pauli-X and Pauli-Z operators. (Right) We represent operators with non-trivial support on region $A$ and $B$ as a graph. Vertex operators are represented as vertices, and plaquette operators are represented as edges. Every edge incident to a vertex represents a plaquette that anti commutes with a star operator. Clearly, the natural generating set is not in canonical form.}
\end{figure}

We now consider the entropy of a ball in the 3D toric code, see FIG.\ref{Jelly}. To the left of this Figure, we show the corner of a region, where the region is filled with transparent green `jelly'. We show some examples of the restriction of star and plaquette operators outside the green jelly. We seek a canonical generating set.

\begin{figure}[b]
\includegraphics{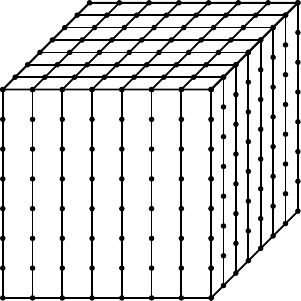}
\caption{The graph for a cuboid-like, ball shaped region in the bulk of the 3D toric code with double edges removed. The top face differs from the side faces due to the anisotropic generating set. The entanglement entropy does not depend on the choice of the generating set. This will become apparent as we progress through the calculation. \label{Cuboid}}
\end{figure}

We simplify FIG.\ref{Jelly} by representing the restricted stabilizers on a graph of vertices and edges. Vertices are denoted by a single index, $a$, and edges take the index of two vertices $(a,b)$ where $ a \not= b$ and $(a,b) = (b,a)$. We show the graph that corresponds to the corner of the region to the right of FIG.\ref{Jelly}. In this graph, vertices represent the restriction of star operators on region $A$, and each edge represents the restriction of an independent plaquette operator. An edge that is incident to a vertex represents a restricted plaquette operator that anti commutes with the respective restricted star operator that is represented by the adjacent vertex. The graph is not in canonical form as there are many edges incident to each vertex.

We show the full graph for the restriction of a ball-shaped region in FIG.\ref{Cuboid}, where any double edges connecting two vertices are removed. We will see why we are free to replace double edges with single edges shortly.

\begin{figure}[t]
\includegraphics{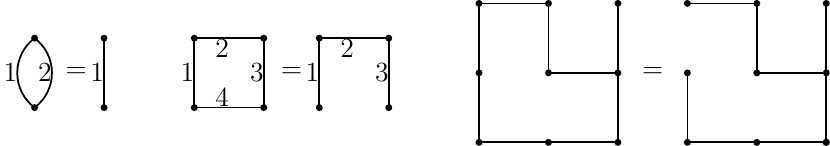}
\caption{(Left) Double edges can be replaced by single edges, without any contribution to the entanglement of the region. (Middle) We replace restricted generator edge 4 with the product of all edges 1, 2, 3 and 4, allowing us to remove edge 4 from the graph. (Right) In general, we can always remove an edge from a circuit due to the circuit rule. \label{Rules}}
\end{figure}

\begin{figure}[b]
\includegraphics{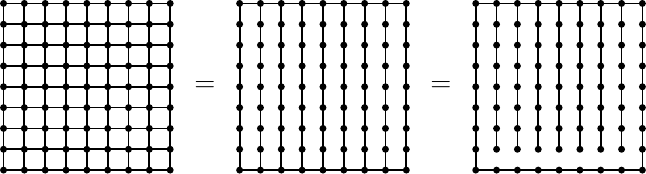}
\caption{We use the rules we have introduced to show that an face with a square grid of edges is equivalent to a face which contains only vertical edges. The right equality is obtained with further use of the circuit rule. \label{Faces}}
\end{figure}

We face the task of finding edges that we are allowed to remove from the graph to find a canonical generating set while still generating $\mathcal{S}$. We complete the entropy calculation by introducing rules that enable us to find canonical form and count the ebits of entanglement shared between the region and its complement.

For a ball-shaped region in the bulk of the lattice, $R$, we recover the known result
\begin{equation}
S(\rho_R) = A_R - 1, \label{Eqn:BulkEntropyResult}
\end{equation} where $A_R$ is the number of star operators with nontrivial support on both subsystem $A$ and subsystem $B$. This is equal to the number of vertices in the graph. We will observe that all but one vertex operator will contribute to the entanglement which gives the result obtained in the literature given our definition of surface area.

We now look to find a canonical generating set. In the first step, we remove double edges, as we have already done in FIG.\ref{Cuboid}. We are free to do this due to the {\em circuit rule}. Before introducing the circuit rule, we first define a {\em series}, and a {\em circuit} of edges.

\begin{defi}[Series] A series of length $x$ is a set of edges $e_j = (a_j, b_j)$ for $1 \le j \le x$ such that $b_j = a_{j+1}$ for $ 1 \le j \le x-1 $. Moreover, each vertex appears in no more than two edges of the series.
\end{defi}

We also define a circuit, which is a special case of a series of edges

\begin{defi}[Circuit]
A circuit is a series of $x$ edges $e_j = (a_j, b_j)$ such that $b_x = a_1$.
\end{defi}

Having introduced a series and a circuit, we are able to introduce the circuit rule

\begin{myrule}[The Circuit rule]
We can remove a single edge from a circuit without affecting the entanglement of the partition.
\end{myrule}

We give examples of circuits and the circuit rule in FIG.\ref{Rules}. To show the circuit rule, we consider the explicit examples of the (Left) and (Middle) cases of circuits shown in FIG.\ref{Rules}. For (Left), we see two edges, $e_1$ and $e_2$. Their corresponding restricted generators anti-commute with the star operators represented by the two vertices adjacent to $e_1$ and $e_2$. To remove the generator of corresponding to $e_2$, we replace it with the product of the generators represented by $e_1$ and $e_2$. This effectively removes $e_2$, as the new restricted generator commutes with all the star operators shown on the graph.

Similarly, as shown in FIG.\ref{Rules}(Middle), we can remove a single edge from four edges, $e_1$, $e_2$, $e_3$ and $e_4$, bounding a square face. We replace the plaquette operator represented by $e_4$ with the product of all the stabilizer generators corresponding to the edges bounding the square such that the new restricted generator commutes with all the vertices of the graph. We thus effectively remove the edge from the graph. This rule trivially generalizes to any circuit. We show this generalization in FIG.\ref{Rules}(Right).

\begin{figure}[t]
\includegraphics{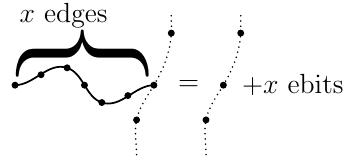}
\caption{An loose end of $x$ black edges equates to $x$ ebits of entanglement. \label{OpenEnd}}
\end{figure}

We apply the circuit rule to the different faces of the graph shown in FIG.\ref{Cuboid}. Using this rule, we obtain the equality shown in FIG.\ref{Faces} between different faces of the cuboid. We reduce all the faces of the cuboid to the form of the right of equality FIG.\ref{Faces} for the next step in the calculation. The new face we find in FIG.\ref{Faces} has {\em loose ends} in the graph

\begin{defi}[Loose end]
A loose end is a series of $x$ edges $e_j = (a_j, b_j)$ such that the only edge incident to vertex $a_1$ is $e_1$, and the only edges incident to vertices $b_j$ are $e_j$ and $e_{j+1}$ for $1\le j \le x-1$.
\end{defi}

The length of a loose end is proportional to its entropy contribution by the loose-end rule

\begin{figure}[b]
\includegraphics{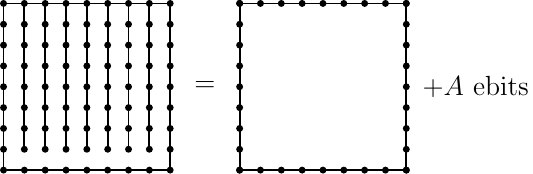}
\caption{The entanglement of a face of the cuboid graph. Each of the $A$ internal vertices on the left hand side of the equality that are removed from the graph in the right hand side of the equality correspond to $A$ ebits of entanglement due to the loose-end rule. \label{FaceEnt}}
\end{figure}

\begin{myrule}[The loose-end rule]
A loose end of $x$ edges denotes $x$ ebits of entanglement shared under the bipartition.
\end{myrule}
We show this rule pictorially in FIG.\ref{OpenEnd}. We see the loose-end rule rigorously by enumerating the restricted stabilizers, $S_j$, along the loose end. Here $S_j$ for odd $j$ are restricted $A_v$ operators, represented vertices in the graph, and restricted $B_f$ operators, edges, have even $j$. The indices take values $1 \le j \le 2x+1$, and $2x+1$ indexes the operator corresponding to the black vertex at the end of the blue string. We have that $\{ S^A_j, \, S^A_{j+1} \} = 0$ for  $1 \le j \le 2x$. We find a canonical form for the edges and vertices of the loose end by making the replacement $S_j \rightarrow S_j' = \prod_{\text{odd}k \le j} S_k$ for all odd $j$, and $S_j \rightarrow S_j' = S_j$ for even $j$. With this replacement we have $ \{ S_{2j-1}'^A, S_{2j}'^A \} = 0 $ for $1 \le j \le x$. We thus identify $x$ ebits of entanglement for a loose end of length $x$.

\begin{figure}[t]
\includegraphics{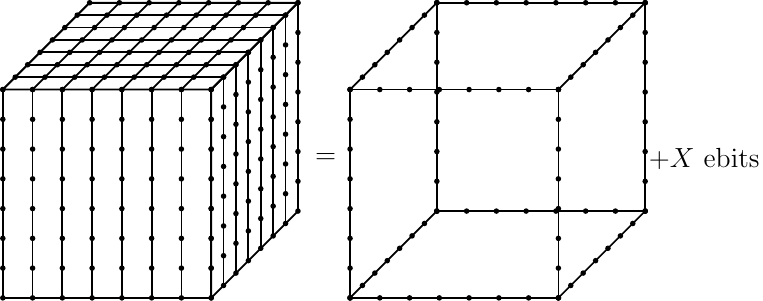}
\caption{We use the repeatedly use the circuit rule and the loose-end rule to find the entanglement represented by the faces of a cuboid graph, where $X$ is the number of vertices removed from all the faces of the cuboid on the right hand side of the equality. \label{CubeEntAndFrame}}
\end{figure}

We now identify the entanglement of a face of a cuboid, as shown in FIG.\ref{FaceEnt}. We use the loose-end rule to see that all the vertices in each face of the graph contribute a single unit of entanglement to the calculation, and thus that all the vertices contribute to the area term of the entropy. We extend this to all the faces of the cube, as shown in FIG.\ref{CubeEntAndFrame}.

\begin{figure}[b]
\includegraphics{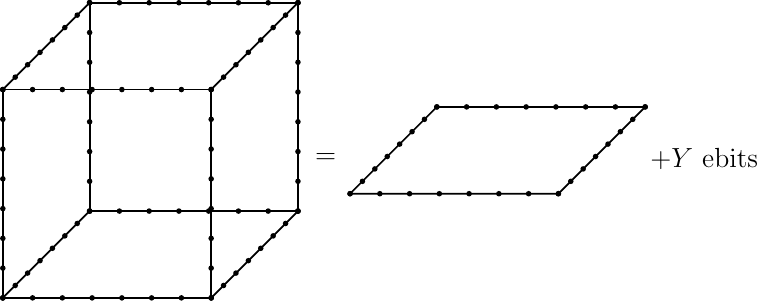}
\caption{\label{OneLoopLeft} We remove $Y$ vertices from the graph on the left hand side of the equation using the circuit rule and the loose-end rule. Each vertex removed contributes a single unit of entanglement in the calculation, we thus identify $Y$ ebits of entanglement on the right hand side of the equality.}
\end{figure}

We can continue to use the circuit rule and the loose-end rule to arrive at the result of FIG.\ref{OneLoopLeft} where only a single loop of edges remains in the graph. Importantly, all the vertices that have been removed from the graph have contributed one unit to the entanglement entropy. To complete the calculation we must assess the entanglement of the single loop of edges that remains in the graph.

\begin{defi}[Loop]
A loop of length $x$ is a circuit of $x$ edges, $e_j = (a_j, b_j)$, such that the only edges of the graph incident to vertices $b_j$ are $e_{j}$ and $e_{j+1}$ for all $j$ where edge $e_{x+1} = e_{1}$.
\end{defi}

Given the definition of a loop, we are now able to introduce the loop rule

\begin{myrule}[The loop rule]
A loop of length $x$ denotes $x-1$ ebits of entanglement under the bipartition.
\end{myrule}

\begin{figure}[t]
\includegraphics{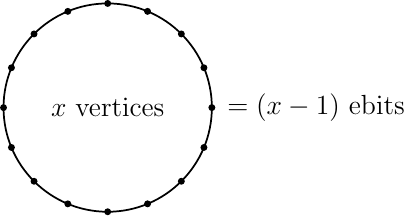}
\caption{\label{ClosedLoopRule} A loop length $x$ denotes $x-1$ ebits of entanglement.}
\end{figure}

We show the loop rule graphically in FIG.\ref{ClosedLoopRule}. We consider the case of a loop carefully. Like the loose end rule, we denote the restricted stabilizer generators in the loop by operators as $S^A_j$, where the indices take values $1 \le j \le 2x$, with even $j$ restricted plaquette operators, edges, and odd $j$ restricted star operators, vertices. Initially, we have that every $S^A_j$ anti commutes with two other restricted generators, $S^A_{j-1}$ and $S^A_{j+1}$, where $S_{2x+1} = S_1$ to accommodate the periodic structure of the loop.

To obtain canonical form, we must replace a single star operator with the product of all the star operators in the loop $S_1 \rightarrow S_1' = \prod_{\text{odd}k} S_k$, such that ${S_1'}^A$ commutes with all the $B_f^A$ denoted by edges in the loop. Similarly, we make the substitution $S_2 \rightarrow S_2' = \prod_{\text{even}k} S_k$, such that ${S_2'}^A$ commutes with all the other restricted operators in the loop. Having removed one edge and one vertex from the loop, we can reduce the remaining $S_j$ with $j > 2$ into canonical form using the loose-end rule. We thus identify $x-1$ ebits of entanglement.

The loop rule removes a single vertex from the graph without contributing to the entanglement, thus giving the universal topological contribution in the calculation, Eq.\ref{Eqn:BulkEntropyResult}.

\begin{figure}[b]
\includegraphics{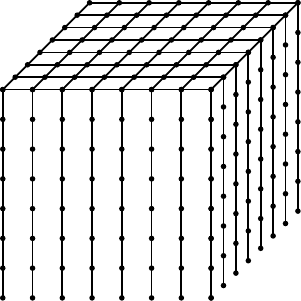}
\caption{We show a graph of the restricted generators for a ball-shaped region pressed against an smooth surface, where the smooth surface is at the bottom of the cuboid. We find the result $S(\rho_R) = A_R - 1$ as in the case in the bulk. \label{SmoothSurfaceBall}}
\end{figure}

One can check that this method extends to any region with a connected boundary, such as an annulus. Ultimately, the calculation will always reduce the graph to a loop. We are then able to remove a single vertex without contributing to the entanglement, thus always giving the desired result for a connected boundary. In general, for regions that include multiple disjoint boundaries, every connected boundary can be reduced to a single loop, enabling us to remove one vertex of the graph per connected boundary without contributing to the entanglement. The topological correction will therefore scale with the number of connected boundaries that enclose the region.

\subsection{A ball on a smooth surface}

\begin{figure}[t]
\includegraphics{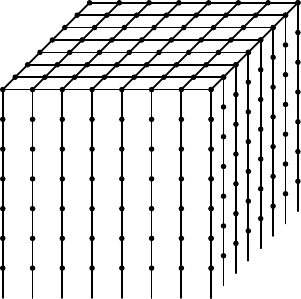}
\caption{The graph for the restricted stabilizers for a region touching a rough face. The bottom edge terminates with edges, not with vertices, so we cannot use the loose end rule to measure the entanglement here. \label{RoughFaceBall}}
\end{figure}

\begin{figure}[b]
\includegraphics{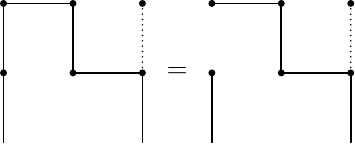}
\caption{ \label{RoughEndCircuit} The graph on the left side of the equality shows an extended circuit of black edges. Both ends of the series terminates at an edge, not a vertex. The right hand side of the equality shows a single edge removed from the extended circuit, as is permitted by the extended circuit rule. }
\end{figure}

\begin{figure}[t]
\includegraphics{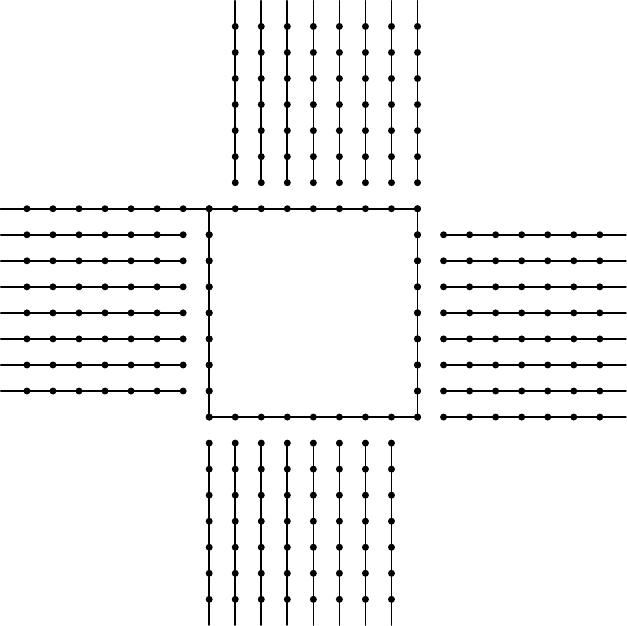}
\caption{\label{OneLegRoughFace} We repeatedly apply the extended circuit rule to detach loose ends from graph. Each of the vertices of these loose ends contribute a single unit of entanglement to the calculation by the loose-end rule.}
\end{figure}

The entanglement entropy of a ball-shaped region that includes qubits from a smooth boundary gives the same topological contribution as the case we previously considered in the bulk. We show the graph of restricted generators of such a region in FIG.\ref{SmoothSurfaceBall}. The result $S(\rho_R) = A_R- 1$ is obtained using the rules we have already established, where $A_R$ is the number of star operators cut by the boundary. Star operators cut near the surface are not treated differently from those cut in the bulk.

\subsection{A ball on a rough surface}

We now consider the entanglement entropy for the case where the region touches the rough face. Contrary to the cases we have considered previously, we do not find a topological contribution to the entanglement entropy. As such, we describe this calculation in detail. We show a picture of the restricted stabilizer graph in FIG.\ref{RoughFaceBall}. Unlike the previous graphs we have considered, here we have edges that have only one adjacent vertex. We denote such an edge as $e_j = (a_j)$, where $a_j$ is the single vertex to which edge $e_j$ is incident. These edges represent restricted plaquette operators that anti commute with only one restricted star operator, denoted by their single incident vertex in the graph. Such plaquette operators are found at the rough boundary of the 3D toric code.

To calculate the entanglement of this region, we introduce the extended circuit rule. We first define an {\em extended circuit}

\begin{defi}[Extended circuit]
An extended circuit is a series of $x$ edges $e_j = (a_j, b_j)$ for $ 2 \le j \le x-1$ and where edges $e_1 =  (b_1)$ and $e_x = (a_x)$ contain a single vertex.
\end{defi}

We are thus able to give the {\em extended-circuit rule} we require to complete the calculation

\begin{myrule}[The extended-circuit rule]
We can remove a single edge from an extended circuit without modifying the entanglement shared across the bipartition.
\end{myrule}

We show the extended circuit rule in FIG.\ref{RoughEndCircuit}. The restriced plaquette operator represented by the missing edge on the right hand side of this equality has been replaced by the product of all restricted plaquette operators represented by solid edges on the left hand side of the Figure, such that the new generator commutes with all the vertices of the graph.

We implement the extended circuit rule many times, together with the circuit rule and the loose-end rule to arrive at the graph shown in FIG.\ref{OneLegRoughFace}.

\begin{figure}[b]
\includegraphics{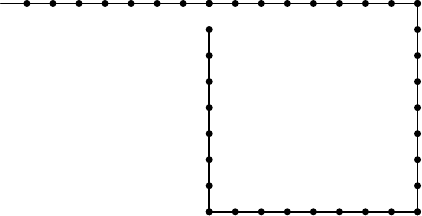}
\caption{\label{LonelyString} The remaining graph of $x$ vertices will contribute $x$ ebits of entanglement due to the loose-end rule.}
\end{figure}

The new graph has many loose ends. Following repeated application of the loose-end rule, and one use the circuit rule gives the graph shown in FIG.\ref{LonelyString}. We complete this entropy calculation with one final application of the loose-end rule. Unlike the previous calculations, all the vertices of the graph have been removed using the loose-end rule, and not once have we obtained a loop which has previously given the topological contribution in the entropy calculation. We thus obtain
\begin{equation}
S(\rho_R) = A_R,
\end{equation}
for the case where $R$ includes qubits of a rough boundary of the 3D toric code.

\bibstyle{plain}

\end{document}